\documentclass{sig-alternate-10pt}
\usepackage[utf8]{inputenc}
\usepackage[utf8]{inputenc}
\usepackage[T1]{fontenc}
\usepackage{enumitem}
\setlist{nolistsep}
\usepackage{subcaption}
\usepackage{tabularx}
\usepackage{booktabs}
\usepackage{multirow}
\usepackage{fancyvrb}
\usepackage{relsize}
\usepackage[bookmarks=false]{hyperref}
\title{Missing measurements on RIPE Atlas}
\subtitle{\vspace{-5mm}A view from probe connection activity}

\numberofauthors{1}
\author{
  \alignauthor 
    Wenqin Shao\textsuperscript{*}, 
    Jean-Louis Rougier\textsuperscript{*},
    François Devienne\textsuperscript{\dag} 
    and Mateusz Viste\textsuperscript{\dag} \\
\affaddr{\textsuperscript{*}Telecom ParisTech, \textsuperscript{\dag} Border 6}
}

\begin{document}

\maketitle

\begin{abstract}
We show that it is common to lose some datapoints for measurements scheduled at regular interval on RIPE Atlas. 
The temporal correlation between missing measurements and connection events are 
analyzed, in the pursuit of understanding reasons behind such missings.
To our surprise, a big part of measurements are lost while probes are connected.
\end{abstract}


\section{Introduction}
RIPE Atlas, a public measurement platform, enables various application such as performance monitoring~\cite{latencymon, Rimondini2014}, anomalies detection~\cite{Fontugne2016, Padmanabhan, halo}, peering and IXP measurements~\cite{ixp, routeixp} etc. 
As it continues to gain popularity among network operators and researchers, its measurement quality becomes a natural concern.
It is now known that load have obvious impacts on measurement precision and scheduling~\cite{Holterbach2015a, Bajpai2015}.

We focus on data completeness, another aspect of measurement quality that received less attention so far. Missing measurements can cause various undesired consequences. Apart from widening confidence interval of inference~\cite{Fontugne2016}, it requires in general methodological adaptations, e.g. in spectrum analysis~\cite{Babu2010, Luckie2014, shao2016}, otherwise biased estimation would be expected~\cite{Baraldi2010}.

One obvious reason of missing is that the probe is not running (properly), e.g. power off~\cite{schedule}.
As long as a probe is powered, it tries to maintain a connection to a controller to report measurements and receive assignments. 
Therefore the probe connection activity provides a good indication of the probe availability, and is used in current investigation conducted by RIPE on probe OS stability~\cite{1look, 2look, 3look}.

In order to infer other possible causes, we crossed the measurement timestamps with the moments probe connects to and disconnects from a Atlas controller.
If measurement missing coincides with the probe disconnection, chances are that the probe is dysfunctional during the missing. However, if measurements are lost while the probe is well connected, something `abnormal' should be expected, beyond the known probe OS issue. 

\section{Data collection}
We observed the RIPE Atlas platform for one month, from 2016-06-01 to 2016-07-01 UTC.
All the v3 probes first connected before the beginning date (11613 of them) are considered.
Connection events (id 7000) and built-in Ping measurements to DNS b-root (id 1010), a highly available destination, are collected~\cite{built-in}. 
Controllers and the ping destination are not within the same network.
Controller logs the moments at which probes connects to and disconnects from it.
The built-in ping measurement is scheduled on every probe at 4min interval. 
10800 ping results are thus expected from each probe within the month.
7353 probes, out of the available 11613, had Ping measurements during this period.

\section{Missing at first glance}
\begin{figure}
\centering
\includegraphics[width=0.34\textwidth]{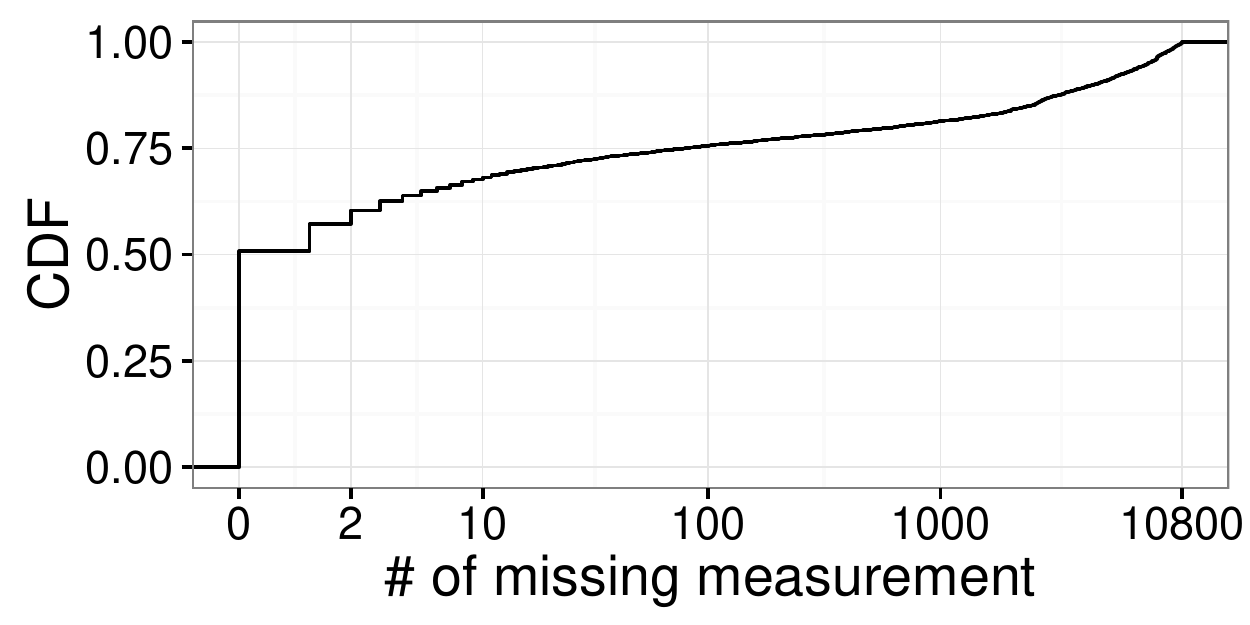}
\caption{CDF of total missing length per probe.}
\label{fig:miss_len}
\end{figure}
4440 probes  (60.4\%) miss no more than 2 datapoints, which is totally legitimate, as random jitter is added to each single measurement to avoid synchronization among probes.
For the rest, the missing length spans a wide range according to Fig.~\ref{fig:miss_len}.
1358 (18.5\%) probes miss more than 10\% of the total measurements (i.e. 72 hours over a month).

\section{Cross with connection events}
Several reasons may contribute to the disconnection of a probe to its controller: 1) probe not working (properly); 2) network issues preventing the connection; 3) controller not available, e.g. during maintenance~\cite{controller}. Meanwhile, the last two reasons shall not prevent a probe from performing built-in measurements, as the results can be unreachable or timeout, and be stored locally on the probe~\cite{usb}. 
That is to say, \textit{missing does not necessarily occur when probe is disconnected, but is unexpected while probe is connected.}

\subsection{Overlap with connected period}
\begin{figure}[!htb]
\centering
\includegraphics[width=0.3\textwidth]{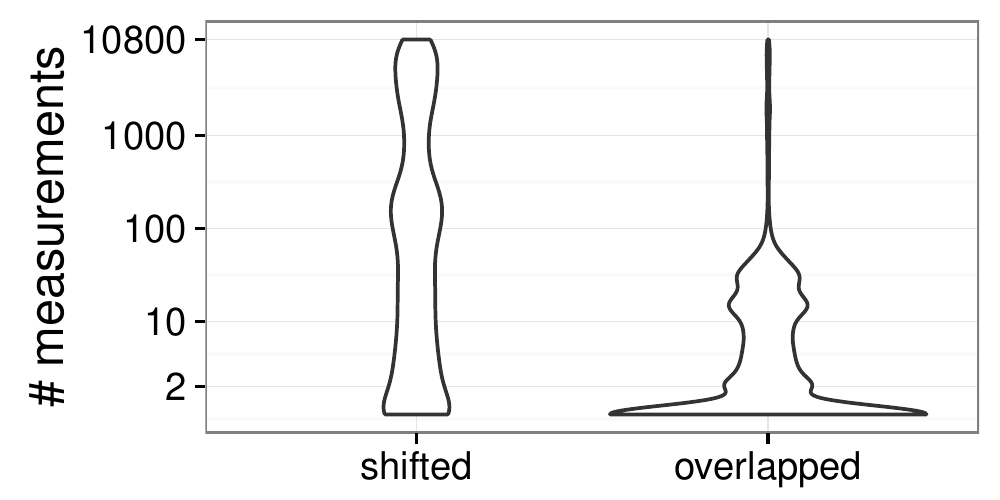}
\caption{Missing length distribution.}
\label{fig:len_ratio}
\end{figure}
We count, for each missing segment, the number of missing measurements that occurred during a connected period.
We obtain the \textit{overlap ratio} by dividing this count by the length of missing segments. 
The distribution of overlap ratio is concentrated at the two ends, 0 and 1. 
For the convenience of illustration, we cut missing segments into two groups, one with overlap ratio $\leq0.5$, denoted as \textit{shifted}, the other with the rest, denoted as \textit{overlapped}.
Measurement missing that overlaps connected period is `unexpected'.

The two groups demonstrate different length distribution profiles, Fig.~\ref{fig:len_ratio}.
15391 missing segments are observed. 
10292 (66.87\%) missing segments are overlapped with connected period. 
They are mostly short in length. 5560 of them last no more than 2 measurements. 
One possible explanation is that these measurements are skipped due to scheduling or load issues~\cite{schedule, Holterbach2015a}.
Meanwhile, 2490 of them are equal to or longer than 1 hour, involving only 620 probes, for which we believe that the previous explanation hardly applies.

Missing segments shifted from connected period are more likely to be long. This is possibly due to the v3 probe OS stability issue still under investigation. It is known to be responsible for long term probe disconnection and requires manual operation to recover the probe~\cite{usb, 1look, 2look, 3look}.

\subsection{Temporal correlation}
\begin{figure}[!htb]
\centering
\includegraphics[width=0.48\textwidth]{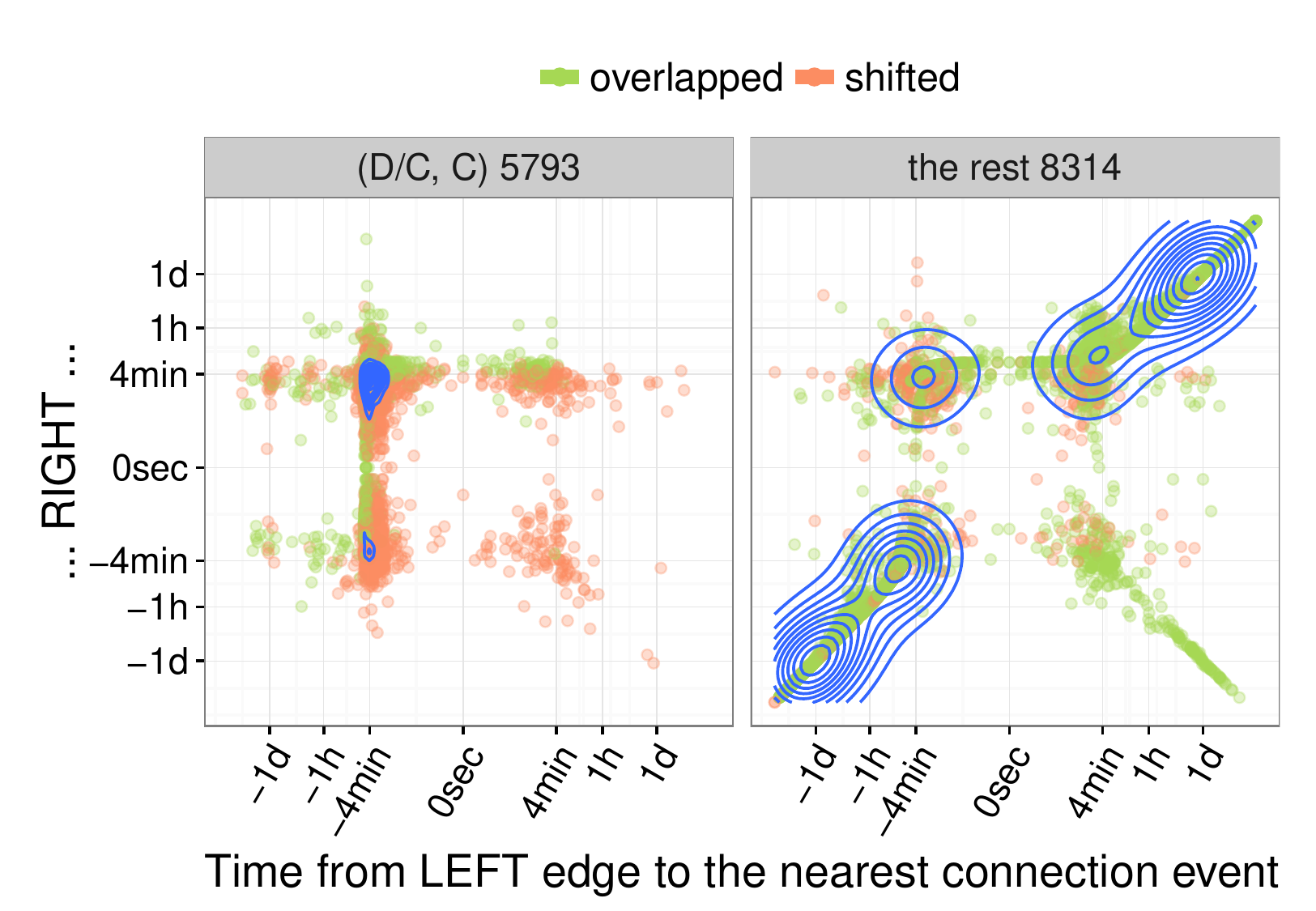}
\caption{(D/C, C) stands for missing segments more closely correlated with disconnected period.
Number of concerned missing segments is given in the title. Negative time distance means the edge happens before the connection event and vice verse.}
\label{fig:all}
\end{figure}
To obtain a subtler view, we seek to find out: when do measurements begin to be lost and when are they recovered? Are these moments close to connection events?

First, we define the \textit{left edge} as the last measurement before a period of measurement absence, and the \textit{right edge}, accordingly, the first measurement after recovery. 
We then calculate the time interval separating these edges and the closest connection events. We also identify the nature of these events, and mark `D/C' for disconnection, `C' for connection. 

1284 missing segments locate at the beginning or the end of the observation period. 
They are unavailable for this analysis as only one edge can be observed.
For the rest with both edges, 5793 missing segments' left edge is closer to a disconnection event and the right edge is closer to a connection event, according to Fig.~\ref{fig:all} sub-graph (D/C, C).
Judging from the density contour, last measurement most likely precedes disconnection by a Ping interval (4min), and recovery tends to take place 4 minute after the connection.
Such strong correlation with probe disconnected period indicates that probe dysfunction is probably the cause.

However, the beginning of such missing segments (D/C, C) can as well be dislocated from disconnection event.
At the left end of the graph, measurements are lost long before the disconnection, which we find `abnormal', even though the recovery is near connection.

To the right end, measurements only begin to be lost a long time after the probe is disconnected. One possible explanation is that measurements are first stored locally after disconnection from controller~\cite{usb}. Then new measurements are lost after local storage is full.

Contrary to the compact distribution in (D/C, C) sub-graph, the majority of the rest missing segments spreads along the time axis. 
The distances from left and right edge to connection events are highly correlated, suggesting that both left and right edges are on the same side of a same connection event.
We note that these measurements were mostly lost while probes were well connected, i.e. overlapped.

\section{Conclusion}
In our analysis covering a large number of probes over one month, only 60\% of v3 Atlas probes have complete measurements. Around 1/3 missing segments appear to closely correlated to disconnected period. The probe OS stability issue might have contributed to such missings, as suggested by the heavy tail of the missing lengths.

However, the remaining 2/3 of missings occurred while probes are connected. 
Half of them are no more than 2 measurements in length, and are thus likely to be caused by scheduling issues. However, around 25\% of this category lasts long($\geq 1h$). The reason of such an unexpected behavior is still to be determined.

\clearpage
\bibliographystyle{abbrv}
\bibliography{ref} 
\end{document}